\documentclass[twocolumn,prb]{revtex4}
\usepackage{amsfonts}
\usepackage[T1]{fontenc}
\usepackage{amsmath,amsbsy,amssymb,graphicx}
\usepackage{times}

\let\mathbf=\boldsymbol

\begin{document}

\title{{\Large Antiferromagnetic Topological Superconductor}\\
{\Large and Electrically Controllable Majorana Fermions}}
\author{Motohiko Ezawa}
\affiliation{Department of Applied Physics, University of Tokyo, Hongo 7-3-1, 113-8656,
Japan }

\begin{abstract}
We investigate the realization of a topological superconductor in a generic
bucked honeycomb system equipped with four types of mass-generating terms,
where the superconductor gap is introduced by attaching the honeycomb system
to an $s$-wave superconductor. Constructing the topological phase diagram,
we show that Majorana modes are formed in the phase boundary. In particular,
we analyze the honeycomb system with antiferromagnetic order in the presence
of perpendicular electric field $E_{z}$. It becomes topological for $%
|E_{z}|>E_{z}^{\text{cr}}$ and trivial for $|E_{z}|<E_{z}^{\text{cr}}$, with 
$E_{z}^{\text{cr}}$ a certain critical field. It is possible to create a
topological spot in a trivial superconductor by controlling applied electric
field. One Majorana zero-energy bound state appears at the phase boundary.
We can arbitrarily control the position of the Majorana fermion by moving
the spot of applied electric field, which will be made possible by a
scanning tunneling microscope probe.
\end{abstract}

\author{}
\maketitle

%\date{}

\address{{\normalsize Department of Applied Physics, University of Tokyo, Hongo
7-3-1, 113-8656, Japan }}

Topological superconductor and Majorana fermion are among the hottest topics
in condensed matter physics\cite{Alicea,Beenakker,Flen,Qi}. Majorana
fermions will be a key player of future quantum computations\cite%
{Ivanov,Nayak}. The anti-particle of a Majorana fermion is itself.
Zero-energy states of a superconductor are necessarily Majorana modes based
on the particle-hole symmetry. One dimensional $p$-wave topological
superconductor\cite{Kitaev,Alicea,Flen,Beenakker} and two-dimensional ($p$+$%
ip$)-wave topological superconductor are fundamental models to realize them%
\cite{Read,Ivanov}. Another promising candidate would be to utilize the
quantum anomalous Hall (QAH) insulator with a proximity-coupled $s$-wave
normal superconductor\cite{QiQAH}. It is a time-reversal breaking
topological superconductor\cite{Schnyder} with class D.

In this paper, we investigate topological superconductivity in a generic
honeycomb system in proximity to an $s$-wave superconductor. Honeycomb
monolayer systems have provided us with an interesting playground of two
dimensional topological insulators. In particular, the buckled honeycomb
system exhibits various topological phases such as the quantum spin Hall
(QSH) insulator, the QAH insulator and the spin-polarized QAH (SP-QAH)
insulator\cite{EzawaPhoto,Ezawa2Ferro} depending on the four mass
parameters. The driving forces are the Kane-Mele spin-orbit interaction\cite%
{KaneMele} with coupling parameter $\lambda _{\text{SO}}$, the staggered
potential with $\lambda _{V}$, the antiferromagnetic order with $\lambda _{%
\text{SX}}$\ and the Haldane term\cite{Kitagawa,EzawaPhoto} with $\lambda _{%
\text{H}}$, generating the mass $\Delta $ to the Dirac fermions intrinsic to
the honeycomb system. The Chern number is calculated to be $\frac{1}{2}$sgn$%
(\Delta )$ for each Dirac cone, which depends on these parameters. The
topological phase diagram is constructed in the ($\lambda _{\text{SO}%
},\lambda _{V},\lambda _{\text{SX}},\lambda _{\text{H}}$) space. There are
important observations. First, it is possible to control the Dirac mass $%
\Delta $ locally by controlling the external forces. The easiest one is the
control of the staggered potential by changing the applied electric field%
\cite{EzawaNJP}. The phase boundary is given by the condition $\Delta =0$.
Hence we are able to accommodate two different topological phases in a
single honeycomb system\cite{EzawaNJP}. Gapless Dirac modes ($\Delta =0$)
are generated along a phase boundary, as is consistent with the bulk-edge
correspondence. They are helical, chiral or spin-polarized modes emerging in
the edge of the QSH, QAH and SP-QAH insulators, respectively.

The honeycomb system is made superconducting in proximity to an $s$-wave
superconductor. The natural question is whether a topological insulator
turns into a topological superconductor. We construct the topological phase
diagram in the ($\lambda _{\text{SO}},\lambda _{V},\lambda _{\text{SX}%
},\lambda _{\text{H}},\Delta _{\text{SC}}$) space, where $\Delta _{\text{SC}%
} $\ is the superconducting gap. Our results read as follows: The
time-reversal symmetry is broken in all topologically non-trivial
superconductors constructed in this system. Gapless Majorana modes emerge in
the phase boundary. Spin-polarized and chiral edge modes yield one and two
Majorana edge modes, respectively. Helical modes are gapped. Consequently,
by controlling external forces locally, we are able to generate Majorana
bound states in the phase boundary of a topological superconductor\ created
within a trivial superconductor sheet.

A special role is played by the SP-QAH insulator\cite{EzawaPhoto,Ezawa2Ferro}
since its Chern number is one. It is an antiferromagnetic topological
insulator. The antiferromagnet order in honeycomb system would be naturally
realized in transition metal oxides\cite{Hu}. We study the antiferromagnetic
topological superconductor obtained by the proximity effect. We apply
electric field locally to the sample. For instance, let us apply it in such
a way that $\lambda _{V}>\lambda _{V}^{\text{cr}}$ for $r<r_{0}$ and $%
\lambda _{V}<\lambda _{V}^{\text{cr}}$ for $r>r_{0}$ in the polar coordinate
with $\lambda _{V}^{\text{cr}}$ being a certain critical potential. One
Majorana fermion is induced at the phase boundary $r=r_{0}$. We may
arbitrarily control its position by moving the region of electric field,
which will be experimentally feasible by a scanning tunneling microscope
(STM) probe.

\textbf{Honeycomb system}: A generic buckled honeycomb system is described
by the four-band tight-binding model\cite{KaneMele,LiuPRB,Ezawa2Ferro}, 
\begin{align}
H_{0}=& -t\sum_{\left\langle i,j\right\rangle \alpha }c_{i\alpha }^{\dagger
}c_{j\alpha }+i\frac{\lambda _{\text{SO}}}{3\sqrt{3}}\sum_{\left\langle
\!\left\langle i,j\right\rangle \!\right\rangle \alpha \beta }\nu
_{ij}c_{i\alpha }^{\dagger }\sigma _{\alpha \beta }^{z}c_{j\beta }  \notag \\
& -\lambda _{V}\sum_{i\alpha }\mu _{i}c_{i\alpha }^{\dagger }c_{i\alpha
}+\lambda _{\text{SX}}\sum_{i\alpha }\mu _{i}c_{i\alpha }^{\dagger }\sigma
_{\alpha \alpha }^{z}c_{i\alpha }  \notag \\
& +i\frac{\lambda _{\text{H}}}{3\sqrt{3}}\sum_{\left\langle \!\left\langle
i,j\right\rangle \!\right\rangle \alpha \beta }\nu _{ij}c_{i\alpha
}^{\dagger }c_{j\beta },  \label{HoneyHamil}
\end{align}%
where $c_{i\alpha }^{\dagger }$ creates an electron with spin polarization $%
\alpha $ at site $i$, and $\left\langle i,j\right\rangle /\left\langle
\!\left\langle i,j\right\rangle \!\right\rangle $ run over all the
nearest/next-nearest neighbor hopping sites. The first term represents the
nearest-neighbor hopping with the transfer energy $t$. The second term
represents the spin-orbit coupling\cite{KaneMele} with $\lambda _{\text{SO}}$%
, where $\nu _{ij}=+1$ if the next-nearest-neighboring hopping is
anticlockwise and $\nu _{ij}=-1$ if it is clockwise with respect to the
positive $z$ axis. The third term is the staggered sublattice potential\cite%
{EzawaNJP} with $\lambda _{V}$, where $\mu _{i}$ takes $1$ ($-1$) for $A$\ ($%
B$) sites. The staggered term may exist intrinsically or induced by applying
electric field $E_{z}$, $\lambda _{V}=\ell E_{z}$. The forth term represents
the antiferromagnetic exchange magnetization\cite{Ezawa2Ferro,Feng} with $%
\lambda _{\text{SX}}$. The fifth term is the Haldane term\cite{Haldane} with 
$\lambda _{\text{H}}$, which will be introduced by applying photo-irradiation%
\cite{Kitagawa,EzawaPhoto}.

\begin{figure}[t]
\centerline{\includegraphics[width=0.5\textwidth]{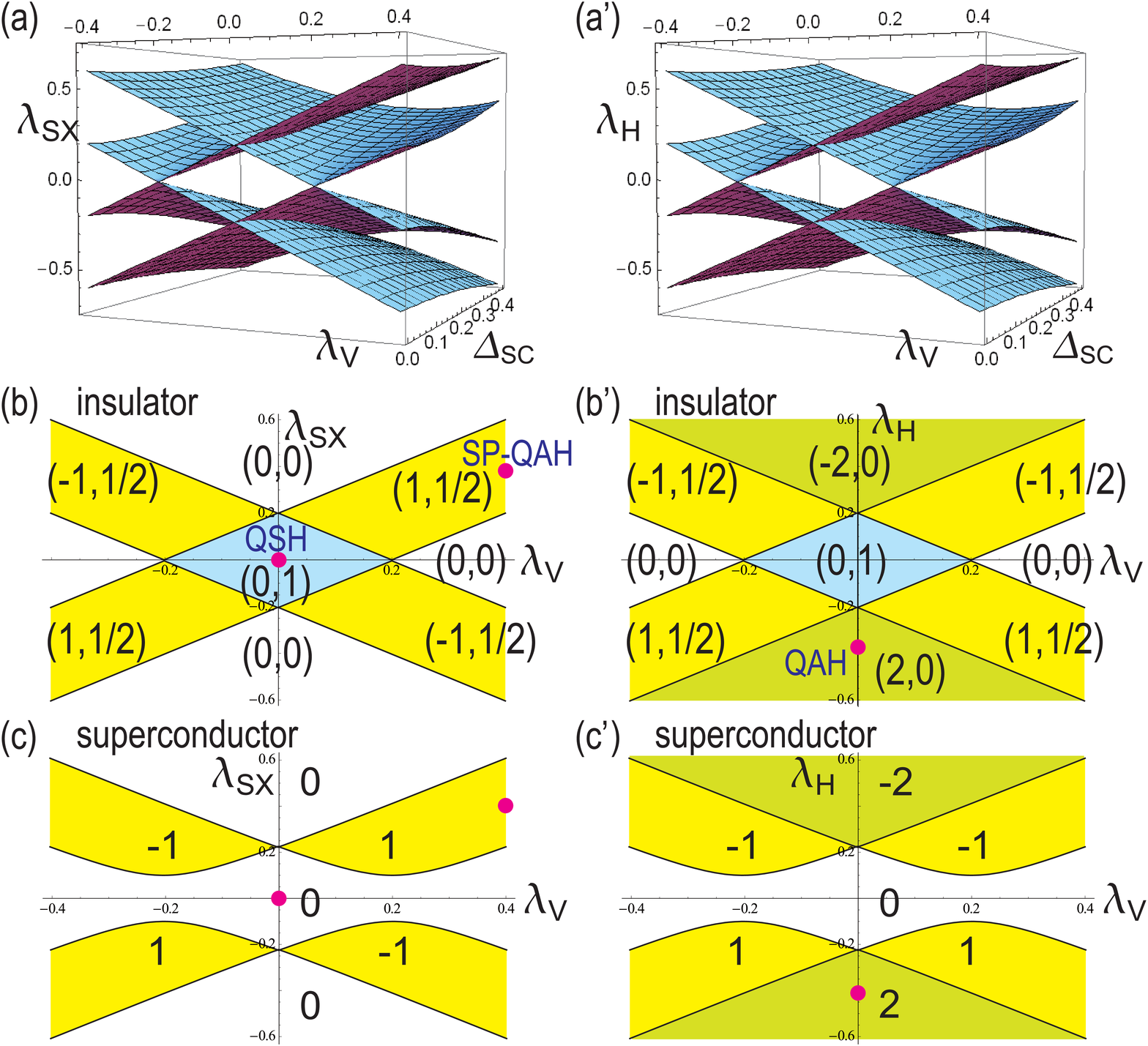}}
\caption{(Color online) Topological phase diagrams in (a) the $(\protect%
\lambda _{V},\protect\lambda _{\text{SX}},\Delta _{\text{SC}})$ space and
(a') the $(\protect\lambda _{V},\protect\lambda _{\text{H}},\Delta _{\text{SC%
}})$ space, where we have taken $\protect\lambda _{\text{SO}}=0.2t$. Phase
diagrams for topological insulator in (b) the $(\protect\lambda _{V},\protect%
\lambda _{\text{SX}})$ space and (b') the $(\protect\lambda _{V},\protect%
\lambda _{\text{H}})$ space. Each phase is indexed by the set $(C,C_{\text{%
spin}})$ of the Chern and spin-Chern numbers. Phase diagrams for topological
superconductor in (c) the $(\protect\lambda _{V},\protect\lambda _{\text{SX}%
})$ space and (c') the $(\protect\lambda _{V},\protect\lambda _{\text{H}})$
space, where we have taken $\Delta _{\text{SC}}=0.1t$. Each phase is indexed
by the Chern number $C$. Tree red dots in (b,b') [(c,c')] show points where
the band structure of nanoribbons are calculated and shown in Fig.\protect
\ref{FigTSC}(a,b,c) [(a',b',c')].}
\label{FigVA}
\end{figure}

\begin{figure}[t]
\centerline{\includegraphics[width=0.5\textwidth]{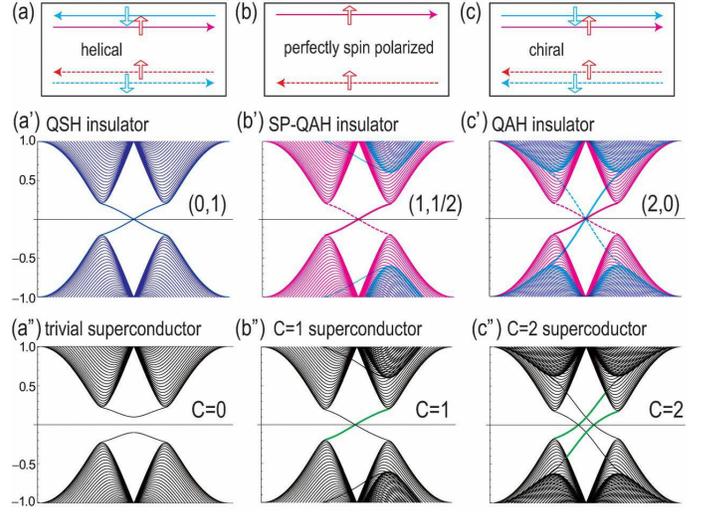}}
\caption{(Color online) Zero-energy edge modes in a nanoribbon for (a) the
QSH insulator, (b) the SP-QAH insulator and (c) the QAH insulator. Up (down)
arrow indicates up (down) spin, while solid (dotted) line indicates right
(left) edge mode. Band structures of these nanoribbons are given in (a'),
(b') and (c'). Majorana modes emerge along a phase boundary between
topological and trivial superconductors. There appear (a") zero, (b") one
and (c") two Majorana modes in superconductor nanoribbons as indicated by
green curves.}
\label{FigTSC}
\end{figure}

\textbf{Topological insulator: }The physics of electrons near the Fermi
energy is described by Dirac electrons near the $K$ and $K^{\prime }$
points, to which we also refer as the $K_{\eta }$ points with $\eta =\pm $.
The effective Dirac Hamiltonian around the $K_{\eta }$ point in the momentum
space reads\cite{EzawaAQHE} 
\begin{align}
H_{\eta }=& \hbar v_{\text{F}}\left( \eta k_{x}\tau _{x}+k_{y}\tau
_{y}\right)  \notag \\
& +\lambda _{\text{SO}}\sigma _{z}\eta \tau _{z}-\lambda _{V}\tau
_{z}+\lambda _{\text{SX}}\sigma _{z}\tau _{z}+\lambda _{\text{H}}\eta \tau
_{z},
\end{align}%
where $\sigma _{a}$ and $\tau _{a}$ are the Pauli matrices of the spin and
the sublattice pseudospin, respectively and $v_{\text{F}}=\frac{\sqrt{3}}{%
2\hbar }at$ is the Fermi velocity. The coefficient of $\tau _{z}$ is the
mass of Dirac fermions in the Hamiltonian, which is composed of four terms,%
\begin{equation}
\Delta _{s_{z}}^{\eta }=\eta s_{z}\lambda _{\text{SO}}-\lambda
_{V}+s_{z}\lambda _{\text{SX}}+\eta \lambda _{\text{H}},  \label{DiracMass}
\end{equation}%
with $s_{z}=\pm $ for the spin direction. The energy spectrum forms a Dirac
cone at the $K_{\eta }$ point with the band gap $2|\Delta _{s_{z}}^{\eta }|$.

The Chern number is obtained for each Dirac cone by calculating the Berry
connection, which is given by $C_{s_{z}}^{\eta }=\frac{1}{2}$sgn$(\Delta
_{s_{z}}^{\eta })$. The total Chern number is $C=C_{\uparrow
}^{+}+C_{\uparrow }^{-}+C_{\uparrow }^{+}+C_{\uparrow }^{-}$, while the
spin-Chern number is $C_{\text{spin}}=C_{\uparrow }^{+}+C_{\uparrow
}^{-}-C_{\downarrow }^{+}-C_{\downarrow }^{-}$. The topological phase
diagram is constructed in the $(\lambda _{\text{SO}},\lambda _{V},\lambda _{%
\text{SX}},\lambda _{\text{H}})$ space by calculating $(C,C_{\text{spin}})$
at each point. The condition of topological insulator is $(C,C_{\text{spin}%
})\neq (0,0)$. In particular, the state with $(0,1)$, $(2,0)$ and $(1,1/2)$
are the QSH, QAH and SP-QAH insulators.

The phase boundary is given by $\Delta _{s_{z}}^{\eta }=0$. We give typical
examples of the phase diagram in the $(\lambda _{V},\lambda _{\text{SX}})$
and $(\lambda _{V},\lambda _{\text{H}})$ spaces in Figs.\ref{FigVA}(b) and
(b'), respectively. The band structures and edge modes are illustrated for
typical topological insulator nanoribbons in Fig.\ref{FigTSC}(a), (b), (c)
and (a'), (b') and (c').

\textbf{Topological superconductor: }A topological superconductor is
obtained from a topological insulator due to the proximity effect\cite%
{FuKane} by attaching an $s$-wave superconductor to it. Indeed, Cooper pairs
are formed\cite{GapClose} between up and down spins at the same site of the
honeycomb system (\ref{HoneyHamil}). The resultant BCS Hamiltonian reads%
\begin{equation}
H_{\text{BCS}}=H_{0}+\sum_{\tau =A,B}\Delta _{\text{SC}}c_{\tau \uparrow
}^{\dagger }\left( i\right) c_{\tau \downarrow }^{\dagger }\left( i\right)
+\Delta _{\text{SC}}^{\ast }c_{\tau \downarrow }\left( i\right) c_{\tau
\uparrow }\left( i\right) ,  \label{SilicBCS}
\end{equation}%
where $H_{0}$ is given by (\ref{HoneyHamil}) and $\Delta _{\text{SC}}$\ is
the superconducting gap. It reads\cite{GapClose} $H_{\text{BCS}%
}=H_{K}+H_{K^{\prime }}+H_{\text{SC}}$ with%
\begin{align}
H_{\text{SC}}=& \sum_{\tau =A,B}[\Delta _{\text{SC}}c_{\tau \uparrow
}^{K\dagger }\left( k\right) c_{\tau \downarrow }^{K^{\prime }\dagger
}\left( -k\right) +\Delta _{\text{SC}}c_{\tau \uparrow }^{K^{\prime }\dagger
}\left( k\right) c_{\tau \downarrow }^{K\dagger }\left( -k\right) ]  \notag
\\
& +\text{h.c.}
\end{align}%
in the momentum representation. A finite gap present in a superconducting
state allows us to evaluate the Chern number of the state to determine
whether it is a topological state. Alternatively we may examine the
emergence of gapless edge modes by calculating the band structure of a
nanoribbon with zigzag edge geometry based on this Hamiltonian. The
emergence of gapless edge modes presents a best signal of a nontrivial
topological structure in the system based on the bulk-edge correspondence:
See Fig.\ref{FigTSC}(a"), (b") and (c").

The BCS Hamiltonian is rewritten into the BdG Hamiltonian,%
\begin{equation}
H_{\text{BdG}}=\left( 
\begin{array}{cc}
H_{K}\left( k\right) & H_{\Delta } \\ 
H_{\Delta }^{\dagger } & -H_{K^{\prime }}^{\ast }\left( -k\right)%
\end{array}%
\right) ,  \label{BdG1}
\end{equation}%
by introducing the Nambu representation for the basis vector, i.e., $\Psi
=\left\{ \psi _{A\uparrow }^{K},\psi _{B\uparrow }^{K},\psi _{A\downarrow
}^{K},\psi _{B\downarrow }^{K},\psi _{A\uparrow }^{K^{\prime }\dagger },\psi
_{B\uparrow }^{K^{\prime }\dagger },\psi _{A\downarrow }^{K^{\prime }\dagger
},\psi _{B\downarrow }^{K^{\prime }\dagger }\right\} ^{t}$.

Diagonalizing the BdG Hamiltonian, we obtain the energy spectrum. It
consists of eight levels with the eigenvalues%
\begin{equation}
E_{\text{BdG}}^{\alpha ,\beta }\left( k\right) =\pm \sqrt{\left( \hbar v_{%
\text{F}}k\right) ^{2}+\left( E_{0}^{\alpha ,\beta }\right) ^{2}}
\label{EnergA}
\end{equation}%
with%
\begin{equation}
E_{0}^{\alpha ,\beta }=\sqrt{\left( \left( \lambda _{\text{SO}}-\alpha
\lambda _{V}\right) ^{2}+\Delta _{\text{SC}}^{2}\right) }+\beta \left(
\lambda _{\text{H}}+\alpha \lambda _{\text{SX}}\right) ,  \label{EqC}
\end{equation}%
where $\alpha $ and $\beta $ takes $\pm 1$. The gap closes ($E_{0}^{\alpha
,\beta }=0$) at%
\begin{equation}
\left( \lambda _{\text{H}}+\alpha \lambda _{\text{SX}}\right) ^{2}=\left(
\lambda _{\text{SO}}-\alpha \lambda _{V}\right) ^{2}+\Delta _{\text{SC}}^{2}.
\label{boundary}
\end{equation}%
Though the original Hamiltonian is an $8\times 8$ matrix, we may decompose
it into 4 independent $2\times 2$ Hamiltonians by the following procedure:
First we diagonalize the Hamiltonian at the $K$ and $K^{\prime }$ points by
the unitary matrix $U$, $U^{-1}H_{\text{BdG}}\left( 0\right) U=$diag.$%
\left\{ E_{\text{BdG}}^{\alpha ,\beta }\left( 0\right) \right\} $. Then we
calculate $U^{-1}H_{\text{BdG}}\left( k\right) U$. The resultant matrix is
constituted of the 4 blocks of $2\times 2$ matrix. As a result,
corresponding to $\alpha ,\beta =\pm 1$, we obtain four sets of the 2-band
theories, 
\begin{equation}
H_{U}\left( k\right) =\left( 
\begin{array}{cc}
\beta E_{0}^{\alpha ,\beta } & \hbar v_{\text{F}}k_{-} \\ 
\hbar v_{\text{F}}k_{+} & -\beta E_{0}^{\alpha ,\beta }%
\end{array}%
\right) .  \label{BdG}
\end{equation}%
This Hamiltonian reproduces the energy spectrum (\ref{EnergA}). We may
interpret $\beta E_{0}^{\alpha ,\beta }$ as the modified Dirac mass due to
the BCS condensation.

It is straightforward to calculate the Chern number of the superconducting
honeycomb system $H_{\text{BdG}}$. It is determined by the sign of the
modified Dirac mass\cite{Diamag},%
\begin{equation}
C=\frac{1}{2}\sum_{\alpha ,\beta =\pm 1}\text{sgn}\left( \beta E_{0}^{\alpha
,\beta }\right) .  \label{Chern}
\end{equation}%
The condition of the emergence of a topological superconductivity is $C\neq
0 $. Note that it is zero when the time-reversal symmetry is present. In
order to obtain a non-zero Chern number, $\lambda _{\text{SX}}$ or $\lambda
_{\text{H}}$ must be nonzero. It should be noticed that the spin-Chern
number is no longer defined due to the BCS condensation of the up-spin and
down-spin electrons.

\begin{figure}[t]
\centerline{\includegraphics[width=0.45\textwidth]{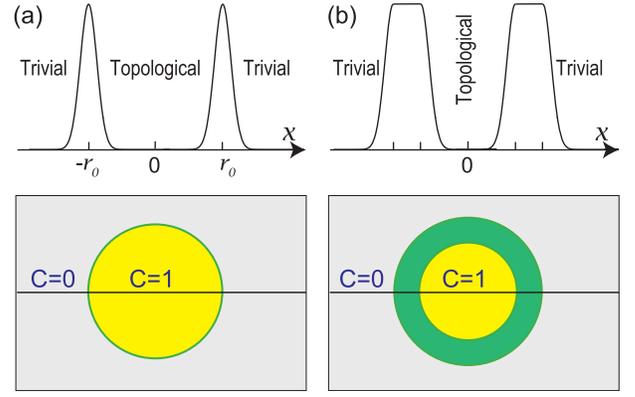}}
\caption{(Color online) Illustration of a Majorana zero-energy state
emerging in antiferromagnetic superconductor. By applying electric field $%
E_{z}$ locally, we may create a topological spot ($C=1$) in a trivial
superconductor ($C=0$). There appears one zero-energy Majorana state in the
phase boundary. (a) The phase boundary is one dimensional in the case of (%
\protect\ref{EqA}). (b) It may be two dimensional when we set $E_{0}^{%
\protect\alpha ,\protect\beta }(r)=0$ for a finite domain of $r$.}
\label{pond}
\end{figure}

The topological phase diagram is easily constructed in the $(\lambda _{\text{%
SO}},\lambda _{V},\lambda _{\text{SX}},\lambda _{\text{H}},\Delta _{\text{SC}%
})$ space. The phase boundaries are given by (\ref{boundary}). The Chern
number is determined from (\ref{Chern}). We show typical examples of the
phase diagram in the $(\lambda _{V},\lambda _{\text{SX}})$ and $(\lambda
_{V},\lambda _{\text{H}})$ spaces in Fig.\ref{FigVA}(c) and (c'). We also
show the band structure of a nanoribbons at a typical point in each phase in
Fig.\ref{FigTSC}(a"), (b") and (c"): There appear no, one and two
zero-energy edge modes, respectively.

\textbf{Zero-energy Majorana bound states: }The zero-energy edge modes are
Majorana bound states due to the electron-hole symmetry. For definiteness we
consider a disk region in a honeycomb sheet, as illustrated in Fig.\ref{pond}%
. We may tune parameters $\lambda _{\text{SO}}$, $\lambda _{V}$, $\lambda _{%
\text{SX}}$, $\lambda _{\text{H}}$ and $\Delta _{\text{SC}}$ to become
space-dependent so that the inner region has a different Chern number from
the outer region. There appears gapless edge modes at the phase boundary.
The Majorana bound states are obtained analytically by solving the BdG
equation.

We take the polar coordinate $\left( r,\theta \right) $. By inserting $%
k_{\pm }=e^{\pm i\theta }(-i\partial _{r}\pm \frac{1}{r}\partial _{\theta })$
into (\ref{BdG}) and setting $\Psi =\{\phi _{A}\left( r\right) e^{in\theta
},\phi _{B}\left( r\right) e^{i(n+1)\theta }\}^{t}$ for the wave function,
the Hamiltonian is written as%
\begin{equation}
H_{U}=\left( 
\begin{array}{cc}
E_{0}^{\alpha ,\beta }\left( r\right) & \hbar v_{\text{F}}(-i\partial _{r}-%
\frac{i(n+1)}{r}) \\ 
\hbar v_{\text{F}}(-i\partial _{r}+\frac{in}{r}) & -E_{0}^{\alpha ,\beta
}\left( r\right)%
\end{array}%
\right) ,
\end{equation}%
where $E_{0}^{\alpha ,\beta }\left( r\right) $ is the inhomogeneous mass (%
\ref{EqC}) with space-dependent parameters $\lambda _{\text{SO}}$, $\lambda
_{V}$, $\lambda _{\text{SX}}$, $\lambda _{\text{H}}$ and $\Delta _{\text{SC}%
} $. By assuming $\phi _{B}\left( r\right) =\pm i\phi _{A}\left( r\right) $\
for $n=-1/2$, the coupled equation $H_{U}\Psi =0$ can be summarized into one
equation\cite{Rossi}%
\begin{equation}
E_{0}^{\alpha ,\beta }(r)\phi _{A}\left( r\right) \pm \hbar v_{\text{F}%
}(\partial _{r}+\frac{1}{2r})\phi _{A}\left( r\right) =0.
\end{equation}%
It can be explicitly solved as%
\begin{equation}
\phi _{A}^{\alpha ,\beta }\left( r\right) =\frac{c_{1}}{\sqrt{r}}\exp \left[ 
\frac{\mp 1}{\hbar v_{\text{F}}}\int_{0}^{r}E_{0}^{\alpha ,\beta }(r^{\prime
})dr^{\prime }\right] ,  \label{EqB}
\end{equation}%
where $c_{1}$ is the normalization constant. The sign $\mp $ is determined
so as to make the wave function finite in the limit $r\rightarrow \infty $.
The zero-energy solution exists at the boundary where the sign of mass term $%
E_{0}^{\alpha ,\beta }(r)$ changes.

In the vicinity of the gap closing point, we can expand as%
\begin{equation}
E_{0}^{\alpha ,\beta }(r)=c_{2}\left( r-r_{c}\right) ,  \label{EqA}
\end{equation}%
where $c_{2}$ is a constant. Substituting this into (\ref{EqB}), we find%
\begin{equation}
\phi _{A}^{\alpha ,\beta }\left( r\right) =\frac{c_{1}}{\sqrt{r}}\exp \left[ 
\frac{-1}{\hbar v_{\text{F}}}\left\vert c_{2}\right\vert r\left( \frac{r}{2}%
-r_{c}\right) \right] .
\end{equation}%
The wave function is Gaussian, where the peak appears at the gap closing
point $r=r_{c}$: See Fig.\ref{pond}(a).

We have derived the wave function for the zero-energy state which emerges
when the mass term $E_{0}^{\alpha ,\beta }(r)$\ vanishes and changes its
sign in general. The zero-energy states with the particle-hole symmetry are
always Majorana fermions. Hence the wave function (\ref{EqB}) represents the
Majorana state.

\textbf{Antiferromagnetic topological superconductor: }There are several way
to make $E_{0}^{\alpha ,\beta }(r_{c})=0$, since there are four independent
mass parameters $\lambda _{\text{SO}}$, $\lambda _{V}$, $\lambda _{\text{SX}%
} $, $\lambda _{\text{H}}$ and one superconducting gap $\Delta _{\text{SC}}$%
. A simple way is to change only one term with fixing all other four terms.

The simplest examples read as follows: (A) The system with $\lambda _{\text{%
SX}}\neq 0$ and $C\neq 0$ has been called an antiferromagnetic topological
insulator\cite{Ezawa2Ferro,Hu}. The associated superconductor may be called
an antiferromagnetic topological superconductor. The number of Majorana
fermions is one, as in Fig.\ref{FigTSC}(b"). (B) The system with $\lambda _{%
\text{H}}\neq 0$ and $C\neq 0$ has been called a photo-induced topological
insulator\cite{Kitagawa,EzawaPhoto}. The associated superconductor may be
called a photo-induced topological superconductor. The number of Majorana
fermions is two, as in Fig.\ref{FigTSC}(c").

We consider explicitly the case where the electric field is applied only to
a disk region in an antiferromagnetic honeycomb sheet, as shown in Fig.\ref%
{pond}. Very strong electric field can be applied experimentally by an STM\
probe. We assume electric field is strong enough to make the system into an
antiferromagnetic topological insulator in the absence of the $s$-wave
superconductivity. Such a field is explicitly given by%
\begin{equation}
\lambda _{V}(r)=\ell E_{z}(r)=\pm \lambda _{\text{SO}}+\sqrt{\lambda _{\text{%
SX}}^{2}-\Delta _{\text{SC}}^{2}}.  \label{LamdaV}
\end{equation}%
The inner region of the disk has a nontrivial Chern number $C=1$ and becomes
a topological superconductor. On the other hands, the outer region of the
disk have $C=0$ and remains to be the trivial superconductor. As a result,
there emerges one Majorana fermion in the phase boundary.

\textbf{Discussions:} Our main observation reads as follows. We are able to
generate a Majorana bound state in an arbitrary position and control it by
moving the spot of applied electric field in an antiferromagnetic
topological superconductor.

Let us briefly discuss experimental feasibility. A best candidate to
materialize such phenomena would be given by transition metal oxide\cite{Hu}%
, where it is estimated that $t\approx 0.2$eV, $\lambda _{\text{SO}}=7.3$%
meV, $\lambda _{V}=\ell E_{z}$, $\lambda _{\text{SX}}=141$meV for LaCrAgO. A
salient property is that the material contains an intrinsic staggered
exchange effect ($\lambda _{\text{SX}}\neq 0$). It has antiferromagnetic
order yielding Dirac mass. We can control the band structure by applying
electric field thanks to the buckled structure.\ When the electric field is
off ($\lambda _{V}=0$), up-spin and down-spin electrons are degenerate. The
degeneracy is resolved as $\lambda _{V}$ increases, and there appear only
up-spin electrons and holes near the Fermi level both for the $K$ and $%
K^{\prime }$ points. The SP-QAH effect is realized. We make this system
superconducting due to the proximity effect. We have derive the critical
electric field (\ref{LamdaV}) to generate a topological spot in an
antiferromagnetic topological superconductor. Using the above sample
parameters we estimate it as $E_{z}^{\text{cl}}=0.1V$\r{A}$^{-1}$. An STM\
probe produces very strong local electric field with circular geometry,
which reaches even a few times larger than this critical value\cite{Gerhard}%
. Our results will open a way of manipulating a Majorana fermion in terms of
electric field. Furthermore, it is possible to control an STM probe very
precisely.

\label{SecConclusion}

I am very much grateful to N. Nagaosa, Y. Tanaka, M. Sato, S. Hasegawa and
N. Takagi for many helpful discussions on the subject. This work was
supported in part by Grants-in-Aid for Scientific Research from the Ministry
of Education, Science, Sports and Culture No. 25400317.

\end{document}